# High transport current superconductivity in powder-in-tube $Ba_{0.6}K_{0.4}Fe_2As_2$ tapes at 27 tesla

He Huang,[1,2] Chao Yao,[1] Chiheng Dong,[1] Xianping Zhang,[1] Dongliang Wang,[1] Zhe Cheng,[1,2] Jianqi Li,[3] Satoshi Awaji,[4] Yanwei Ma[1,2,a]

[1] *Key Laboratory of Applied Superconductivity, Institute of Electrical Engineering, Chinese Academy of Sciences, Beijing 100190, China*

[2] *University of Chinese Academy of Science, Beijing, 100049, China*

[3] *Beijing National Laboratory for Condensed Matter Physics, Institute of Physics, Chinese Academy of Sciences, Beijing 100190, China*

[4] *High Field Laboratory for Superconducting Materials, Institute for Materials Research, Tohoku University, Sendai 980-8577, Japan*

*E-mail: ywma@mail.iee.ac.cn

[a] Author to whom any correspondence should be addressed.

**Abstract**

The high upper critical field and low anisotropy of iron-based superconductors make them being particularly attractive for high-field applications. However, the current carrying capability needs to be enhanced by overcoming the weak-link effect between misaligned grains inside wire and tape conductors. Here we demonstrate a high transport critical current density ($J_c$) reaching $1.5 \times 10^5$ A/cm$^2$ ($I_c$ = 437 A) at 4.2 K and 10 T in $Ba_{0.6}K_{0.4}Fe_2As_2$ (Ba-122) tapes prepared by a combination of conventional powder-in-tube method and optimized hot-press technique. The transport $J_c$ measured at 4.2 K under high magnetic fields of 27 T is still on the level of $5.5 \times 10^4$ A/cm$^2$, which is much higher than those of low-temperature superconductors. This is the first report of hot-pressed Ba-122 superconducting tapes and these $J_c$ values are by far the highest ever reported for iron-based superconducting wires and tapes. These high-performance tapes exhibit high degree of c-axis texture of Ba-122 grains and low anisotropy of transport $J_c$, showing great potential for construction of high-field superconducting magnets.

**Introduction**

Since the advent of iron-based superconductors in 2008,[1-6] their advantages, including relatively high transition temperature $T_c$ (~38 K),[3] high upper critical field $H_{c2}$ (>100 T),[7,8] superior transport $J_c$ (~10$^6$ A/cm$^2$) and moderate $H_{c2}^{ab}/H_{c2}^{c}$ anisotropy (~2),[9,10] attract many researchers to engage in their mechanism of superconductivity and applications.[11] Although the cuprate superconductors hold the highest $J_c$ value among all superconductors, the obstacles such as the extremely high anisotropies and weak link behavior at grain boundary limit their applications. In addition, the brittle texture property in cuprate superconductor, such as YBCO, imposes a restriction on its fabrication and increases the manufacturing costs. For iron-based superconductors, the inter-grain critical current density $J_c^{gb}$ decrease rapidly when the misalignment of crystalline orientation angle at grain boundaries is larger than 9°[12] and this value is larger than that in YBCO.[13] Currently, the Nb$_3$Sn, a low $T_c$ superconductor is still adopted to generate high field at 4.2 K. However, the sensitivity to strain in Nb$_3$Sn makes this type of superconductor need a post-winding heat-treatment and this process increased the complexity in fabricating wires. We recommend that the iron-based superconductors may be a more competitive candidate in the competition of superconductors for high-field applications.

Significant researches have been undertaken to fabricate iron-based superconducting wires and tapes, among which the powder in tube (PIT) method is widely adopted.[14] However, the inevitable pores and cracks between grains caused by PIT process bring about poor connectivity and weak link behavior, which severely limit the current flow between grains. In the past several years, various strategies were proposed to overcome these obstacles such as metal additions,[15] flat rolling,[16,17] uniaxial pressing,[18,19] and hot isostatic pressing.[20,21] These methods are effective in increasing the core density, texture and connectivity of grains. For example, Sn-doped superconducting powder can enhance the crystallization of grains during the heat treatment,[22] and this enhancement is also effective in "1111" type tapes.[23] The mechanical deformation in the PIT process can significantly enhance the superconducting core density and the uniaxial rolling and press can induce c-axis texture. In addition, reducing voids and inducing oriented c-axis texture, which is crucial to relieve the problem of weak link behavior are the key points to further improve the $J_c$ performance.

Previous works in fabricating tapes with hot-pressing process are usually using Sr-122 materials. The c-axis texture of the rolled Sr-122 superconducting tapes can be significantly increased by the hot-pressing process. In addition, we find the texture value of the flat-rolled Ba-122 tapes are higher than that of Sr-122 tapes,[16,17] which reminds us to investigate the Ba-122 tapes by hot-pressing process. In this work, high-performance $Ba_{0.6}K_{0.4}Fe_2As_2$ superconducting tapes were successfully fabricated by optimized hot-pressing process. The transport $J_c$ at 4.2 K reaches $1.5 \times 10^5$ A/cm$^2$ (10 T) and $5.5 \times 10^4$ A/cm$^2$ (27 T), respectively, and this is the first report in Ba-122 superconducting tapes with such high $J_c$. What is more, these $J_c$ values are the highest values in iron-based superconducting wires and tapes ever reported. The c-axis orientation $F$ value for the hot-pressed (HP) Ba-122 tapes is 0.87 and this value is higher than that of the flat-rolled Ba-122 tapes and the HP Sr-122 tapes.[16,24] We utilize the EBSD technique to further investigate the texture and misorientation angle of the HP tapes and find a high degree of c-axis textured superconducting core which improve the connectivity between grains. With these results, we recommend that the Ba-122 materials are flexible to build textured construction and the high degree textured Ba-122 tapes are beneficial to improve the transport property.

**Experimental details**

The precursor of Ba-122 was prepared by a solid-state reaction method and the tape was prepared by the *ex-situ* PIT process with Sn (5%) as additive. Ba fillings (99%), K pieces (99.95%), As (99.95%) and Fe (99.99%) powders with a nominal composition of $Ba_{0.6}K_{0.5}Fe_2As_2$ were loaded into Nb tube after ball-milling for 10 hours under argon atmosphere. Then the tube was heat-treated at 900 ℃ for 35 hours. The sintered precursor was ground into powder, mixed with Sn material in an agate mortar and finally packed into Ag tube (Outer diameter: 8 mm and inner diameter: 5 mm). Then the Ag tube was swaged and drawn into a wire with diameter of 1.9 mm and flat rolled into tapes. The hot-pressing process was adopted to the final tapes with a pressure about 15 MPa at 880 ℃ for 1 hour.

The transport critical current $I_c$ was measured by the standard four-probe method with a criterion of 1 μV/cm under 4.2 K at the High Field Laboratory for Superconducting Materials (HFLSM) at Sendai. We carried out $I_c$ measurement in a 14 T superconducting magnet and we also measured the $I_c$ properties of the tapes by a hybrid magnet with the fields up to 27 T. The phase construction of superconducting cores was measured by X-ray diffraction (XRD) using Bruker D8 Advance diffractometer with Cu Kα radiation after peeling off the Ag sheathe. The Resistance vs. Temperature curve and the Moment vs. Temperature curves with VSM plugin were measured on Physical Property Measurement System (PPMS). In order to investigate the vortex pinning mechanism of our tape, we used the Dew-Hughes model[25] by the magnetization property of superconducting core with SQUID-VSM on Magnetic Property Measurement System (MPMS, Quantum Design). The crystal orientation was analyzed by EBSD plugin equipped on Scanning Electron Microscope (SEM, Zeiss SIGMA). The details of grain boundaries were examined on a JEOL JEM-2100F TEM.

**Results and discussion**

Fig. 1(a) shows the magnetic field dependence of the transport critical current density $J_c$ at 4.2 K of Ba-122 tapes in this work and the properties of other type of superconducting wires or tapes are also included.[26-29] The transport $J_c$ at 4.2 K in this work reaches $1.5 \times 10^5$ A/cm$^2$ ($I_c$ = 437 A) at 10 T and $5.5 \times 10^4$ A/cm$^2$ at 27 T, respectively. These $J_c$ values are the highest values not only for the Ba-122 tapes but also for all of the iron-based superconducting wires and tapes ever reported. The extrapolated $J_c$ curves of the two batches of samples intersects with the $J_c$ curve of $Nb_3Sn$ around the field of 18 T, indicating that the iron-based superconductor is a competitive candidate for high-field applications. As shown in Fig. 1(b), we also measured the transport $J_c$ with the tape surface parallel, vertical and an angle of 45 ° to the direction of magnetic field, respectively. A group of schematic diagrams were included in the inset of Fig. 1(b) to show the angle between the tape surface and the direction of magnetic field. It can be seen clearly that the $J_c$ values measured with the tape surface vertical to the field direction are higher than that of parallel to the field direction. These results are consistent with the hot-pressed (HP) Sr-122 superconducting tapes.[30] The inset of Fig. 1(b) gives the details of $J_c$ values at 10 T with different magnetic field direction. The anisotropy of the HP Ba-122 tape at 10 T is 1.37 and this value is much smaller than that of the Bi2223 and YBCO. These results demonstrate a nearly isotropic behavior of transport $J_c$ under magnetic fields in different directions, suggesting that the iron-based superconductors are beneficial to manufacture superconducting coils.

The electromagnetic properties of the superconducting core were examined. Fig. 2(a) shows the curves of the resistivity as a function of temperature and Fig. 2(b) exhibits the details. The $T_c^{onset}$ and $T_c^{zero}$ are 37.6 K and 37.3 K at 0 T, respectively. The resistivity transition width $\Delta T_c = 0.3$ K is sharper than that of the Sr-122 tapes by hot-pressing process[18] and this result demonstrates that the electromagnetic properties of the tape are homogeneous. The magnetization as a function of temperature for the HP tape was measured with a 20 Oe magnetic field parallel to the tape surface as shown in Fig. 2(c). The rather sharp superconducting transition indicates high quality superconducting phases and nearly the whole bulk of superconducting core we measured exhibits superconductivity property. Furthermore, the FC curve is independent of temperature indicating strong vortex pinning in the tape.

XRD patterns of the superconducting core of HP tape are shown in Fig. 2(d) and the data of precursor powder is also included for comparison. A well-defined $ThCr_2Si_2$-type crystal structure can be found from the XRD patterns and the Ag peaks are also indexed because of the Ag sheath. Obviously, the (*00l*) peaks were strongly enhanced by the cold-work deformation and the hot-pressing process when comparing the peaks between the superconducting core and the randomly oriented precursor. The degree of c-axis texture can be calculated by the Lotgering method[31] with $F = (\rho - \rho_0)/(1 - \rho_0)$, where $\rho = \Sigma I(00l)/\Sigma I(hkl)$, $\rho_0 = \Sigma I_0(00l)/\Sigma I_0(hkl)$. $I$ and $I_0$ are the intensities of each peak for the textured and randomly oriented samples, respectively. The $F$ value is 0.87 which is higher than that of the hot-pressed Sr-122 tapes and the cold-pressed Ba-122 tapes.[24,32] These results indicate that the crystal orientation of Ba-122 is prone to rotate along the tape surface by the pressure which applied to the tape surface and thus the c-axis texture was enhanced. In addition, the core density is another important parameter which relates to the transport $J_c$ and sometimes shows a positive relation with the transport $J_c$.[32] However, the average Vickers hardness value of the superconducting core of the HP tape is 138, higher than that of the flat-rolled tape but lower than that of the stainless steel sheathed cold-pressed tapes (~200).[16,32] It is obvious that the core density of Ba-122 tapes were enhanced by the hot-pressing process, but the materials of the silver sheath restrict the further enhancement of the superconducting core density. Comparing the texture and core density of the tapes which fabricated by cold-pressing method with our results, we hold that the texture may a dominant parameter which influence the transport $J_c$ of Ba-122 tapes in this work.

The electron backscatter diffraction (EBSD) technique is an important tool to thoroughly characterize the microstructure, grain size and crystal orientation of the superconducting core.[33] The grain orientation and the misorientation angle between grains can be marked by the automatic detected Kikuchi patterns after polishing the center section of superconducting core well. Fig. 3(a) shows the grain size with the neighboring grains marked with different colors in order to identify the different grains. Most grains with the diameter lower than 2 μm are evenly constructed parallel to the c-axis. The grain size of the HP Ba-122 superconducting tape is smaller than that of the HP Sr-122 tapes.[24] In order to quantitatively describe the grain size properties, we plot grain numbers and areas as functions of grain size diameter, which is shown in the inset of Fig. 4. The grain size diameter is defined as $D = 2(A/\pi)^{1/2}$, where A is the area of detected grains. Most grains with diameter of 0.5-1 μm are detected from the EBSD image and this diameter is much smaller than that of HP Sr-122 tapes and the flat-rolled Ba-122 tapes. As we know, in a specific area, the more grains with small size usually means the larger grain boundaries density. Here, the grains with diameter between 1–2.5 μm occupy the main area indicating a homogeneous grain size in the superconducting core.

It is found that grain boundary pinning is dominant in the vortex pinning mechanism in HP Sr-122 superconducting tapes.[34] We plot the $J_c^{mag1/2} \times H^{1/4}$ as a function of H to calculate the irreversibility field $H_{irr}$ as shown in the inset of Fig. 5 where $J_c^{mag}$ is the magnetic critical current density. The $J_c^{mag}$ is obtained from the isothermal hysteresis loops based on the Bean model[35]. We estimate the $H_{irr}$ value by linear fitting the curves to zero $J_c^{1/2} \times H^{1/4}$ value.[36] According to the Dew-Hughes model,[25] if a dominant pinning mechanism exists in the superconducting tape, a unified function $f = Ah^p(1-h)^q$ can be used at different temperature where $f$ is the normalized vortex pinning forces and A is a constant. The results are shown in Fig. 5 in which the curves of $f(h, T)$ at different temperature show the same trend and almost overlap with each other. This result demonstrate that the pinning mechanism is independent of temperature and the function $f = Ah^p(1-h)^q$ can be used in this situation. The cyan color curve in Fig. 5 is fitted to the $f(h, T)$ curves at different temperature with p = 0.64 and q = 2.3. We can calculate the peak position by $h_{max} = p/(p+q) = 0.22$ which is the same with the observed $h_{max}$ value from fitted curve. The $h_{max}$ value is close to 0.2 and according to the Dew-Hughes model, the pinning mechanism of the HP tape belongs to surface pinning.[34] We recommend that the grain boundary pinning is dominant in our tape which is the same with the HP Sr-122, $Nb_3Sn$ and $MgB_2$ superconducting wires and tapes.[37,38] Then the relative small grain size is beneficial to the improvement of the vortex pinning property. This phenomenon also can be found in $MgB_2$ superconductors where the pinning force increases with the decreasing grain size when the sample with good grain connectivity.[39]

Fig. 3(b) shows the inverse pole figure (IPF) map which is given in [001] direction with the color code drawn in the stereographic triangle. The red color reveals that the dominant orientation of grains is (001), indicating the strong c-axis texture and it is also confirmed by the IPF figure shown in Fig. 3(d). These results clarify that the c-axis of almost all of the Ba-122 grains are perpendicular to the tape surface. Fig. 3(c) directly shows the misorientation angle loaded to the grain boundaries. The blue and green color indicate that the misorientation angles between grains at low angle hold a considerable proportion which can be indexed from the color code. We also list the quantified number fraction diagrams of the determined misorientation angle in the range of 1 ° to 90 ° which is shown in Fig. 4. It clarifies that the number fraction, which represents the grain boundary length, decreases with the misorientation angle and about 42.8% of number fraction are within 9 ° which is higher than that of the Sr-122 superconducting tapes.[24] In addition, the average confidence index, the average image quality and the number of indexed good points are also included in the bottom of Fig. 3(d) indicating the EBSD results are reliable enough.

The TEM technology is adopted to further investigate the microstructure of the grain boundaries. Fig 6(a) shows three grains marked with A, B and C with the tape surface direction and the grain boundaries are clean and well-connected. Fig. 6(b) is a typical grain boundary with a misorientation angle 6 ° and many grain boundaries with low misorientation angle can be found from the superconducting core which is confirmed by the results of EBSD. Combining the data of TEM and EBSD, we recommend that the rolling deformation in the PIT process and the force applied to the tape during the heat treatment can effectively enhance the grain connectivity and improve the c-axis texture.

In the PIT process, the c-axis texture produced during the rolling deformation as well as the residual microcracks. However, the hot-pressing process makes these microcracks more flexible to couple with each other and thus inhibit the crumple of grains and continue to improve the c-axis texture. This result is also confirmed by the IPF figures from EBSD. In addition, there are 42.8 % number fraction within the limit of 9 ° in misorientation angle and a large amount of grain boundaries with low angle are detected by the TEM which indicating that the weak-link behavior is suppressed in our tape. Therefore, the strong degree of c-axis texture and the improved connectivity between grains are the reason to the high $J_c$ value.

**Conclusion**

In this work, we systematically studied the properties of the high-performance Ba-122 superconducting tapes fabricated by hot-pressing process. The transport $J_c$ values achieve $1.5 \times 10^5$ A/cm$^2$ ($I_c$ = 437 A) at 10 T and $5.5 \times 10^4$ A/cm$^2$ at 27 T, respectively. This is the first report in Ba-122 tapes with such high transport $J_c$ values and these are also the highest values ever reported in iron-based superconducting wires and tapes. These values are superior to NbTi, Nb$_3$Sn and MgB$_2$ tapes or wires. We find a high degree c-axis texture in the Ba-122 tapes and about 42.8% of number fraction of grain boundaries are within 9 °. These $J_c$ results further strengthen the position of iron-based superconductor as a competitor to other superconductors in high field applications.

**Acknowledgements**

The authors thanks Chen Li for help and useful suggestion. This work is supported by the National Natural Science Foundation of China (Grant Nos. 51320105015 and 51677179), the Beijing Municipal Science and Technology Commission (Grant No. Z171100002017006), the Bureau of Frontier Sciences and Education, Chinese Academy of Sciences (QYZDJ-SSW-JSC026)

**Captions**

FIG. 1. (a) The magnetic field dependence of transport critical current density $J_c$ for the hot-pressed Ba-122 tape at 4.2 K. The transport $J_c$ of other wires or tapes are also included for comparison. (b) The magnetic field dependence of the transport $J_c$ measured with the direction of magnetic field parallel, vertical and 45 ° to the tape surface, respectively. The inset of (b) shows the details of $J_c$ values at 10 T with different magnetic field direction.

FIG. 2. The electromagnetic properties and crystal microstructure of the superconducting core of HP tape. (a) and (b) The resistivity as a function of temperature. (c) Temperature dependence of the susceptibility for the superconducting core with zero field cooling (ZFC) and field cooling (FC) procedures. (d) XRD patterns for the superconducting core of the HP tape. The data of randomly orientated powder is also included as a reference.

FIG. 3. EBSD images for the superconducting core of the HP tape viewed from the ND direction of the tape. (a) The marked neighboring grains with different colors. (b) The inverse pole figure (IPF) image in [001] direction. (c) The misorientation angle loaded to the grain boundaries. The color codes of this image are posted out on the bottom right corner of this image. (d) The inverse pole figure in [001] direction for the measured area.

FIG. 4. Misorientation angle distribution for the superconducting core of the HP tape. The inset shows the grain size distribution compared with the grain number (blue line) and the area (red line), respectively.

FIG. 5. Normalized vortex pinning force $f = F_p/F_p^{max}$ as a function of the reduced field $h = H/H_{irr}$. The fitting curve using the formula $f = Ah^p(1-h)^q$ with cyan color is also included. The inset shows the Kramer plot for 22 K to 32 K with a step length 2 K per curve. The curves are linear fitted with dash line.

FIG. 6. The TEM image of the superconducting core. (a) The clean and well-connected grain boundaries. (b) A random grain boundary with small misorientation angle.

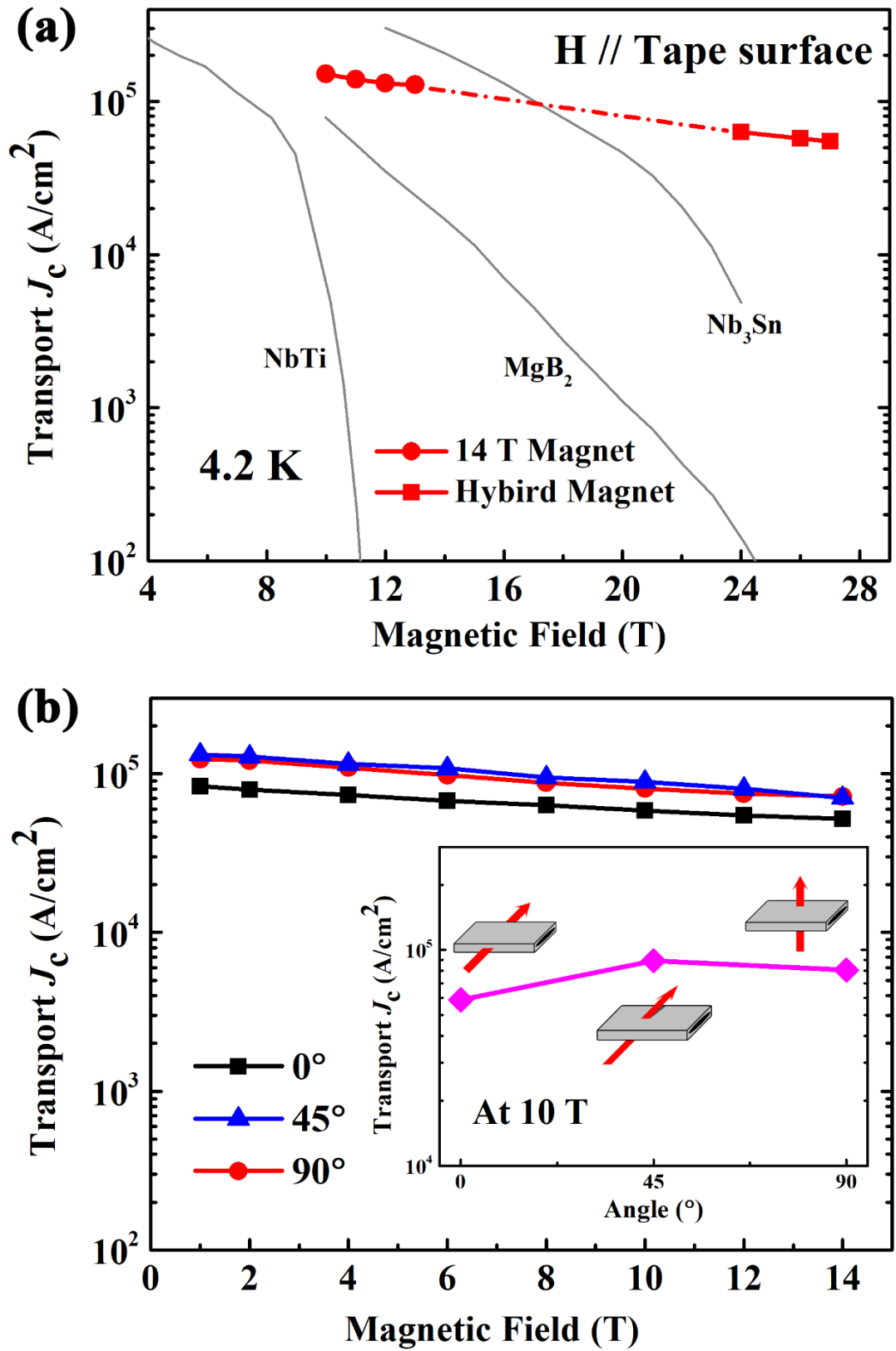

FIG. 1. Huang *et al*

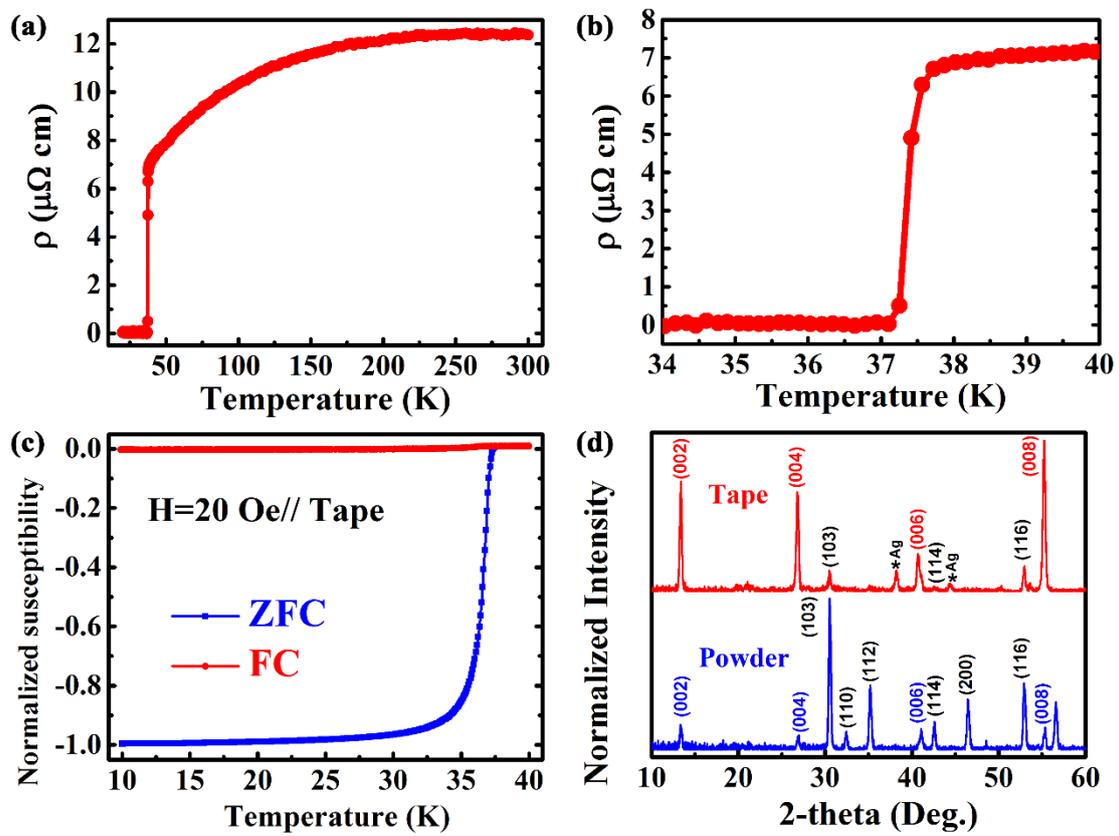

FIG. 2. Huang *et al*

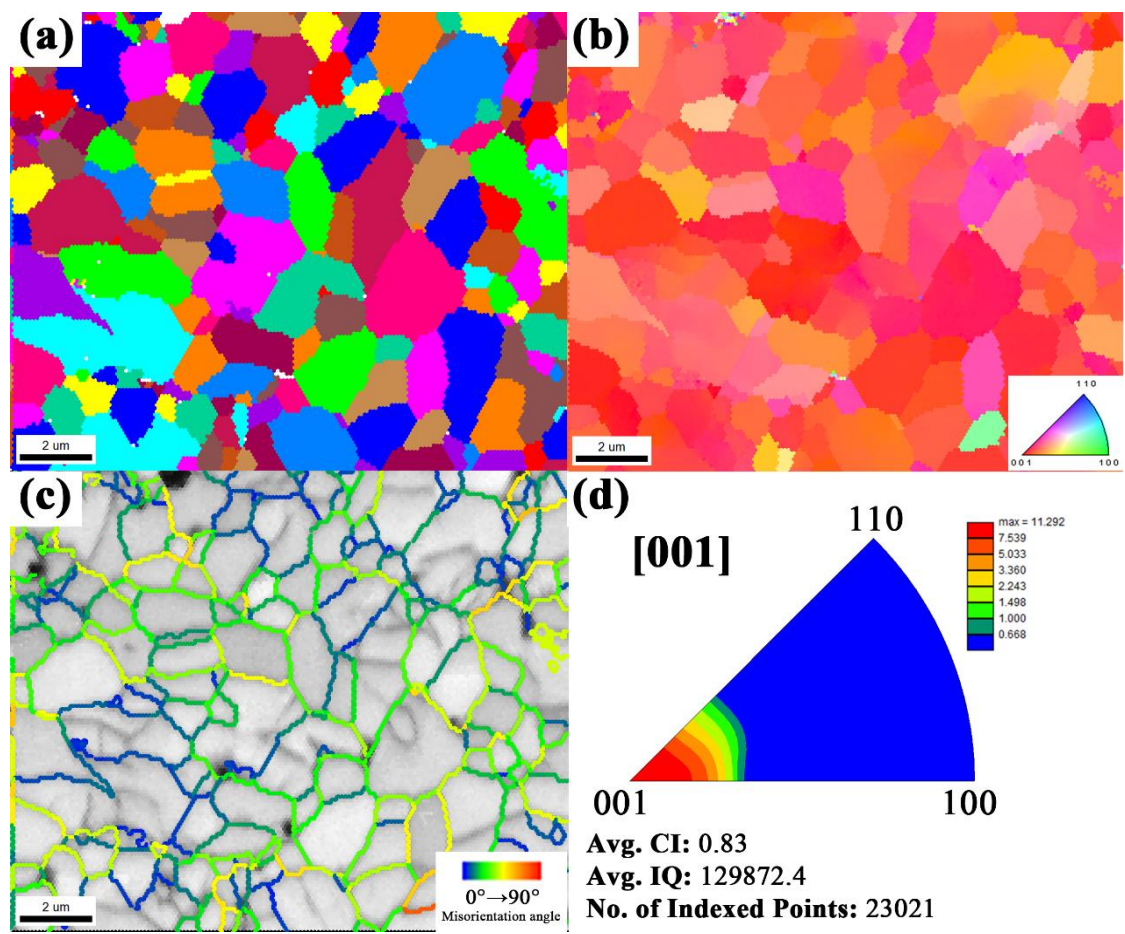

FIG. 3. Huang *et al*

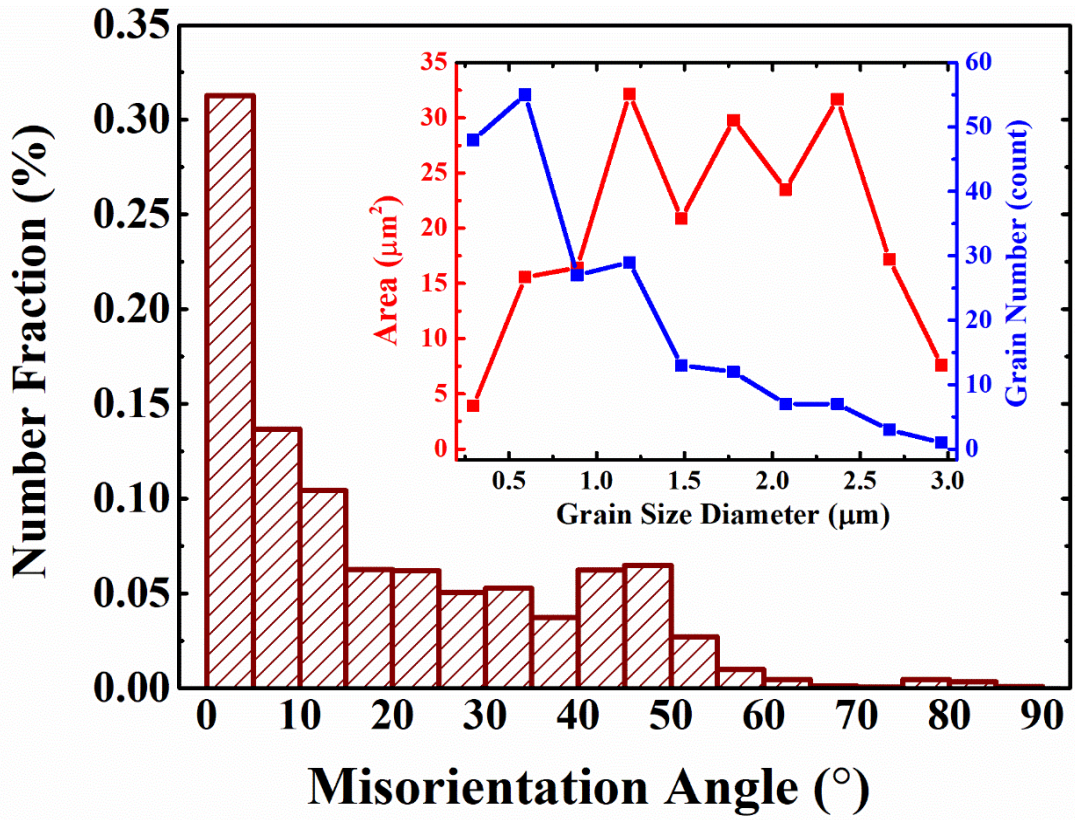

FIG. 4. Huang *et al*

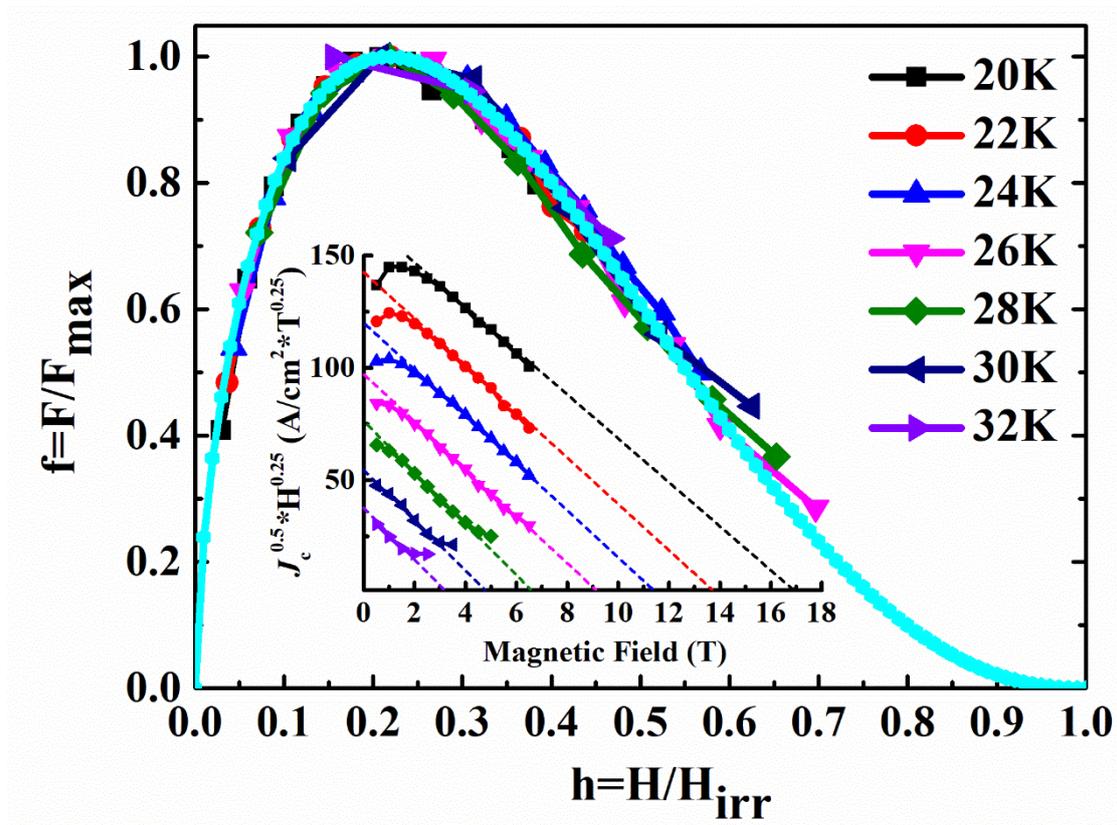

FIG. 5. Huang *et al*

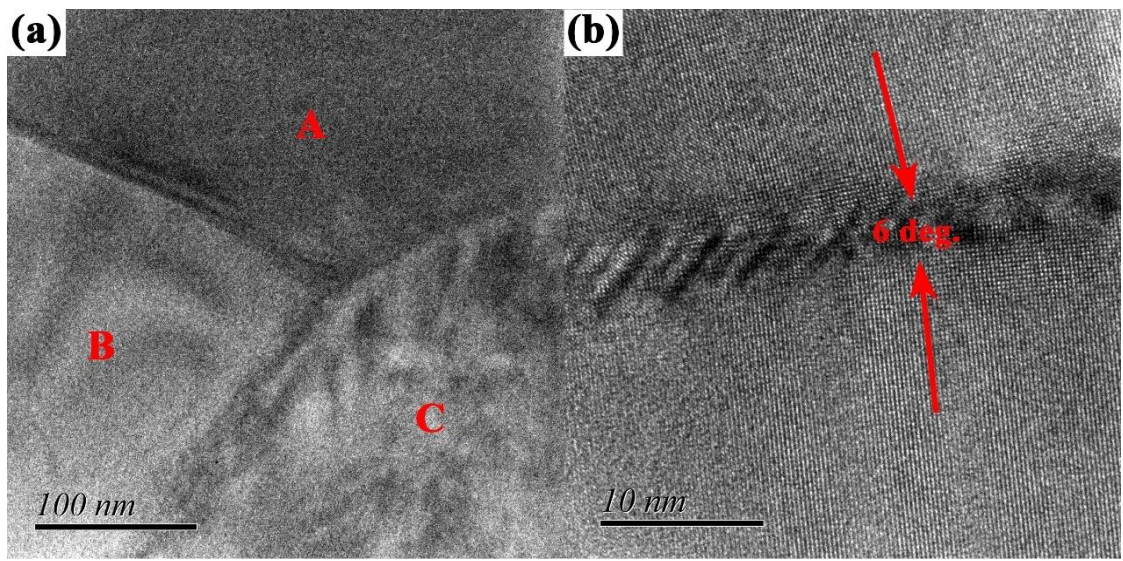

FIG. 6. Huang *et al*